\documentclass[showpacs,amssymb,preprint,preprintnumbers]{revtex4}

\usepackage{amsmath}

\begin{document}

\title{Stress-energy Tensor Correlators in N-dim Hot Flat Spaces via the Generalized Zeta-Function Method}
\author{H. T. Cho}
\email[Email: ]{htcho@mail.tku.edu.tw}
\affiliation{Department of Physics, Tamkang University, Tamsui, New Taipei City, Taiwan}
\author{B. L. Hu}
\email[Email: ]{blhu@umd.edu}
\affiliation{Maryland Center for Fundamental Physics,  University of Maryland, College Park, Maryland 20742-4111, USA \\ and \\Institute for Advanced Studies and Department of Physics, \\ Hong Kong University of Science and Technology, Clear Water Bay, Kowloon, Hong Kong}

\centerline{ \it Dedicated to Professor J. Stuart Dowker on the occasion of his 75th birthday}

\begin{abstract}
We calculate the expectation values of the stress-energy bitensor defined at two different spacetime points $x, x'$ of a massless, minimally coupled scalar field with respect to a quantum state at finite temperature  $T$  in a flat $N$-dimensional spacetime  by means of the generalized zeta-function method. These correlators, also known as the noise kernels, give the fluctuations of energy and momentum density of a quantum field which are essential for the investigation of the physical effects of negative energy density in certain spacetimes or quantum states.  They also act as the sources of the Einstein-Langevin equations in stochastic gravity which one can solve for the dynamics of metric fluctuations as in spacetime foams. In terms of constitutions these correlators are one rung above (in the sense of the correlation -- BBGKY or Schwinger-Dyson -- hierarchies) the mean (vacuum and thermal expectation) values of the stress-energy tensor which drive the  semiclassical Einstein equation in semiclassical gravity.  The low and the high temperature expansions of these correlators are also given here:  At low temperatures, the leading order temperature dependence goes like $T^{N}$  while at high temperatures they have a $T^{2}$ dependence with the subleading terms exponentially suppressed by $e^{-T}$. We also discuss the singular behaviors of the correlators in the $x'\rightarrow x$ coincident limit as was done before for massless conformal quantum fields.
\end{abstract}

\date{February 3, 2012}
\maketitle
\section{Introduction}

In this paper we present a calculation of the stress tensor correlators, also known as the noise kernels, of a  massless, minimally coupled scalar field at a finite temperature  $T$  in a flat $N$-dimensional spacetime  by means of the generalized zeta-function method. The stress tensor correlator is the expectation value of the stress-energy bitensor defined at two separate spacetime points. The zeta-function method was first introduced by Dowker and Critchley \cite{DowCri76} and Hawking \cite{Haw77} (see also \cite{zeta}) and successfully applied to the regularization of ultraviolet divergences in stress tensors of quantum fields in curved spacetimes with Euclidean sections, e.g., static black holes and (anti-)de Sitter space. It was generalized by Phillips and Hu \cite{PH97} (see also \cite{CogEli02}) for the calculation of stress energy correlators. In an earlier paper \cite{ChoHuAdS} we mentioned three classes of physical problems of current interest which necessitate the knowledge of such quantities and motivated us to undertake this task for AdS spaces. Here we also mention three classes of problems as motivation for our present endeavor of calculating the stress energy correlators for finite temperature quantum fields in curved spacetimes. Common to both endeavors foremost is

A) {\it the  semiclassical stochastic gravity program} \cite{stograv}, a theoretical framework established in the 90s as a natural extension of semiclassical gravity (SCG)
theory of the 80s \cite{HuVerSSG}, for including the effects of fluctuations in the quantum matter field through
the Einstein-Langevin equation (ELEq) which govern the behavior of the induced metric fluctuations
\cite{HRV}. While SCG goes beyond quantum field theory in curved spacetime (QFTCST) of the 70s \cite{BirDav} (viewed as the test-field approximation of SCG on a fixed background spacetime), in that the backreaction of the quantum matter
field on the dynamics of the spacetime is incorporated through the expectation value of the stress-energy tensor
as the source of the semiclassical Einstein (SCEq) equation, stochastic gravity goes beyond SCG in that it includes also the backreaction of the fluctuations of the stress-energy tensor, measured by  the expectation values of the stress-energy bitensor, also known as the noise kernel, which govern the behavior of the induced metric fluctuations.

B)  {\it Black Hole fluctuations and backreaction}: Owing to the existence of an Euclidean section in static  (e.g., Schwarzschild) black hole spacetimes quantum field effects can be obtained via thermal field methods. E.g., Hawking effect can be derived with the use of the Euclidean Green function \cite{HarHaw76} whose periodicity corresponds to the inverse Hawking temperature.  Fluctuations and backreaction of the stress-energy tensor near the black hole event horizon is an important issue \cite{HuRou07}, the investigation of this problem for Schwarzschild black holes requires a knowledge of the thermal stress tensor correlator, which is the expectation value of the stress energy bitensor with respect to the Hartle-Hawking state. Thermal stress tensor correlator calculation in hot flat space is the logical first step towards this goal, as was the motivation in the earlier work of Phillips and Hu \cite{PH03}. In that paper the authors use the Gaussian approximation \cite{Gaussian} for the Green function for such quantum fields to evaluate the noise kernel in two optical metrics: hot flat space and  the optical-Schwarzschild spacetime. The optical metric for an ultrastatic spacetime has the product form $ ds^2= g_{ab}dx^a dx^b= d\tau^2 + g_{ij}dx^i dx^j $.
In the Euclidean sector we can allow the imaginary $\tau$ time to possess a periodicity of $2\pi/ \kappa = \beta = 1/T$ , with T the temperature, to connect with thermal field theories. (For a black hole, $\kappa$ is the surface gravity.) For massless conformally coupled quantum fields in hot flat space, the Gaussian Green function is exact. For optical Schwarzschild, the Gaussian Green function is known to be a fairly good approximation for calculating the stress tensor which involves second covariant derivatives of the Green function. The noise kernels involve up to four covariant derivatives of the Green function \cite{PH01}.  The Schwarzschild metric is obtained from the optical Schwarzschild by a conformal transformation.

This earlier work is followed by a recent paper \cite{EBRAH} wherein the authors, instead of seeking the coincident limits of the noise kernel, calculated the point-separated expression (in both timelike and spacelike directions)  necessary to solve the equations of stochastic semiclassical gravity, by the same Gaussian approximation for the Wightman function for conformally invariant fields. These authors have computed all components of the noise kernel exactly for hot flat space and several components  for the Schwarzschild spacetime. They showed that the noise kernel for the conformally invariant field has a simple scaling behavior under conformal transformations which enables them to obtain the results for Schwarzschild spacetime from that for the optical Schwarzschild metric.

C) {\it Negative energy density and fluctuations}  Apart from applications to early universe and black hole physics, closer to home, the  stress energy correlators are directly relevant to the fluctuations of vacuum energy density, especially the existence of \textit{negative} energy density as in flat space with boundaries  (such as the Casimir energy) and for nonclassical states (such as two-mode squeezed states). This issue was raised by Kuo and Ford \cite{KuoFor94}, furthered by Wu and Ford  \cite{WuFor} and pursued by Phillips and Hu \cite{PH97,PH00}, where the  fluctuations in energy density are shown to be comparable to its mean value for several classes of spacetimes (e.g., Einstein Universe) and states (e.g., Casimir). Many related problems such as the quantum interest principle \cite{ForRom} bear on foundational issues of quantum field theory and spacetime structure.  With generalization to finite temperature one can ask how these quantum vacuum behaviors are altered by thermal fluctuations. Results for hot flat space were obtained in \cite{PH03} for conformally invariant quantum fields. Our present results for massless minimally coupled fields in arbitrary dimensions complement and further these earlier studies.

The organization of this paper is as follows: In the next section we introduce the generalized zeta-function method, in Sec III we use this method to calculate the stress energy tensor correlator for a finite temperature quantum field in flat N-dim space. In Sec. IV we give the expressions for high and low temperature expansions showing which components in the bitensor are dominant and how temperature and dimensionality alter this behavior.  We also discuss the singular behaviors of the correlators in the $x'\rightarrow x$ coincident limit. We conclude with some remarks in Sec. V. The detailed expressions are collected in Appendices A, B and C. Our convention is the same as in our earlier papers \cite{ChoHuAdS}.

\section{Generalized Zeta-Function Method}

In this section we first introduce the  generalized zeta-function method of Phillips and Hu \cite{PH97} (see also \cite{CogEli02}) based on the original methods of Dowker and Critchley \cite{DowCri76} and Hawking \cite{Haw77}. This method has recently been used to consider the case of the stress-energy correlations in AdS spaces \cite{ChoHuAdS}. Here we apply it to calculate such correlations in hot flat spaces.

To be general we start by considering a massive $m$ scalar field $\phi$ coupled to an $N$-dimensional Euclideanized space (with contravariant metric $g^{\mu\nu}(x)$, determinant $g$ and scalar curvature $R$) with coupling constant $\xi$ described by the action
\begin{eqnarray}
S[\phi]=\frac{1}{2}\int d^{N}x\sqrt{g(x)}\phi(x)H\phi(x),
\end{eqnarray}
where $H$ is the quadratic operator
\begin{eqnarray}
H=-\Box+m^{2}+\xi R,  \label{H}
\end{eqnarray}
and $R$ is the scalar curvature. The effective action defined by $W={\rm ln}{\cal Z}$ is related to
the generating functional ${\cal Z}$ by 
\begin{eqnarray}
{\cal Z}=\int{\cal D}\phi\ \! e^{-S[\phi]}.
\end{eqnarray}
The expectation value of the stress-energy tensor can be obtained by taking the functional derivative of the effective action
\begin{eqnarray}
\langle T_{\mu\nu}\rangle=-\frac{2}{\sqrt{g(x)}}\frac{\delta W}{\delta g^{\mu\nu}(x)}.
\end{eqnarray}
This formal expression is divergent at coincident limit and some procedure of regularization need be implemented. Here we adopt the $\zeta$-function regularization scheme of  \cite{DowCri76,Haw77}.

The $\zeta$-function of an operator $H$ is defined as
\begin{eqnarray}
\zeta_{H}(s)=\sum_{n}\left(\frac{\mu}{\lambda_{n}}\right)^{s}={\rm Tr}\left(\frac{\mu}{H}\right)^{s}
\end{eqnarray}
where $\lambda_{n}$ are the eigenvalues of $H$ and $\mu$ represents the renormalization scale.
The $\zeta$-function regularized effective action of the operator $H$ is
\begin{eqnarray}
W_{R}=\left.\frac{1}{2}\frac{d\zeta}{ds}\right|_{s\rightarrow 0}.
\end{eqnarray}
Using the proper-time method of Dowker and Critchley \cite{DowCri76}  one can write the $\zeta$-function as
\begin{eqnarray}
\zeta_{H}(s)&=&\frac{\mu^{s}}{\Gamma(s)}\int_{0}^{\infty}dt\ \! t^{s-1}{\rm Tr}e^{-tH}\\
W_{R}&=&\frac{1}{2}\frac{d}{ds}\left[\frac{\mu^{s}}{\Gamma(s)}\int_{0}^{\infty}dt\ \! t^{s-1}{\rm Tr}e^{-tH}\right]_{s\rightarrow 0}
\end{eqnarray}
Taking the first variation of the $\zeta$-function,
\begin{eqnarray}
\delta\zeta_{H}&=&-\frac{\mu^{s}}{\Gamma(s)}\int_{0}^{\infty}dt\ \! t^{s}{\rm Tr}\left(\delta He^{-tH}\right)\nonumber\\
&=&-\frac{\mu^{s}}{\Gamma(s)}\int_{0}^{\infty}dt\ \! t^{s}\sum_{n}e^{-t\lambda_{n}}\langle n\left|\delta H\right|n\rangle
\end{eqnarray}
we obtain the regularized expectation value of the stress-energy tensor given by
\begin{eqnarray}
\langle T_{\mu\nu}(x)\rangle&=&\frac{1}{2}\frac{d}{ds}\left[-\frac{\mu^{s}}{\Gamma(s)}\int_{0}^{\infty}dt\ \! t^{s}\sum_{n}e^{-t\lambda_{n}}\left(-\frac{2}{\sqrt{g(x)}}\left\langle n\right|\frac{\delta H}{\delta g^{\mu\nu}(x)}\left|n\right\rangle\right)\right]_{s\rightarrow 0}\nonumber\\
&=&-\frac{1}{2}\frac{d}{ds}\left\{\frac{\mu^{s}}{\Gamma(s)}\int_{0}^{\infty}dt\ \! t^{s}\sum_{n}e^{-t\lambda_{n}}\, T_{\mu\nu}\left[\phi_{n}(x),\phi_{n}^{*}(x)\right]\right\}_{s\rightarrow 0}\label{renstress}
\end{eqnarray}
where
\begin{eqnarray}
T_{\mu\nu}\left[\phi_{n}(x),\phi_{n'}^{*}(x)\right]&\equiv&-\frac{2}{\sqrt{g(x)}}\left\langle n'\right|\frac{\delta H}{\delta g^{\mu\nu}(x)}\left|n\right\rangle\nonumber\\
&=&-\frac{2}{\sqrt{g(x)}}\int d^{N}x'\sqrt{g(x')}\phi_{n'}^{*}(x')\left[\frac{\delta H}{\delta g^{\mu\nu}(x)}\phi_{n}(x')\right]
\end{eqnarray}
Here $\phi_{n}(x)$ are eigenfunctions  of the operator $H$ corresponding to the eigenvalues $ \lambda_n $, namely,
\begin{equation}
H \phi_n = \lambda_n \phi_n.
\end{equation}
With the form of $H$ given by (\ref{H}) we have
\begin{eqnarray}
T_{\mu\nu}\left[\phi_{n}(x),\phi_{n'}^{*}(x)\right]&=&
-\left(\partial_{\mu}\phi_{n'}^{*}\partial_{\nu}\phi_{n}+\partial_{\nu}\phi_{n'}^{*}\partial_{\mu}\phi_{n}\right)
+g_{\mu\nu}\left(g^{\alpha\beta}\partial_{\alpha}\phi_{n'}^{*}\partial_{\beta}\phi_{n}+\phi_{n'}^{*}\Box\phi_{n}\right)\nonumber\\
&&\ \ \ -2\xi\left[g_{\mu\nu}\Box(\phi_{n'}^{*}\phi_{n})-\nabla_{\mu}\nabla_{\nu}(\phi_{n'}^{*}\phi_{n})+R_{\mu\nu}\phi_{n'}^{*}\phi_{n}\right]
\end{eqnarray}

Using the Schwinger method \cite{Sch51}, Phillips and Hu \cite{PH97} generalized the $\zeta$-function method for the  consideration of the stress-energy correlators. The second variation of the $\zeta$-function can be written as
\begin{eqnarray}
\delta_{2}\delta_{1}\zeta_{H}&=&\frac{\mu^{s}}{2\Gamma(s)}\int_{0}^{\infty}du\int_{0}^{\infty}dv(u+v)^{s}(uv)^{\nu}\nonumber\\
&&\ \ \ \ \ \left\{{\rm Tr}\left[(\delta_{1}H) e^{-uH}(\delta_{2}H) e^{-vH}\right]+{\rm Tr}\left[(\delta_{2}H) e^{-uH}(\delta_{1}H) e^{-vH}\right]\right\}
\end{eqnarray}
and
\begin{eqnarray}
\Delta T_{\mu\nu\alpha'\beta'}^{2}(x,x')&=&\frac{1}{2}\frac{d}{ds}\left\{\frac{\mu^{s}}{\Gamma(s)}
\int_{0}^{\infty}du\int_{0}^{\infty}dv(u+v)^{s}(uv)^{\nu}\sum_{n,n'}e^{-u\lambda_{n}-v\lambda_{n'}}\right.\nonumber\\
&&\ \ \ \ \ \ \left. T_{\mu\nu}\left[\phi_{n}(x),\phi_{n'}^{*}(x)\right]T_{\alpha'\beta'}\left[\phi_{n'}(x'),\phi_{n}^{*}(x')\right]\right.\!{\Bigg \}}_{s,\nu\rightarrow 0}.\label{T2exp}
\end{eqnarray}
Note that in this prescription  an additional regularization factor $(uv)^{\nu}$ has been introduced.
This is because the authors of \cite{PH97} were interested in the fluctuations of the stress-energy tensor, that is, in the coincident limit of $\Delta T_{\mu\nu\alpha'\beta'}^{2}(x,x')$ where under this limit further divergences occur which call for an additional regularization factor. (See also \cite{CogEli02}). However,  our present purpose is focused on getting the correlators with two points separated, i.e., in the non-coincident case. Hence, apart from the fact that the expression in Eq.~(\ref{T2exp}) is more symmetric with this factor, the keeping of this factor above is actually a matter of convenience. Here we can first take the $s\rightarrow 0$ limit without spoiling the regularization and the expression in Eq.~(\ref{T2exp}) becomes
\begin{eqnarray}
\Delta T_{\mu\nu\alpha'\beta'}^{2}(x,x')&=&\frac{1}{2}
\int_{0}^{\infty}du\int_{0}^{\infty}dv (uv)^{\nu}\sum_{n,n'}e^{-u\lambda_{n}-v\lambda_{n'}}\nonumber\\
&&\ \ \ \ \ \ T_{\mu\nu}\left[\phi_{n}(x),\phi_{n'}^{*}(x)\right]T_{\alpha'\beta'}
\left[\phi_{n'}(x'),\phi_{n}^{*}(x')\right]\!{\bigg |}_{\nu\rightarrow 0}.\label{T2finalexp}
\end{eqnarray}
We shall see from the following calculations that with this expression the integrations over $u$ and $v$ effectively separate. The calculations are therefore simplified considerably.

\section{Stress energy correlators in hot flat spaces}

We now apply this generalized $\zeta$-function method to calculate the stress energy correlators of massless, minimally coupled quantum fields in hot flat spaces. The N-dimensional Euclidean flat space metric has the form
\begin{equation}
ds^{2}=d\tau^{2}+d\vec{x}\cdot d\vec{x}
\end{equation}
where $\tau$ has periodicity $\beta=1/T$. The operator $H$ is just $-\Box$ and the corresponding normalized eigenfunctions are
\begin{equation}
\phi_{n\vec{k}}(\tau,\vec{x})=\frac{1}{(2\pi)^{(N-1)/2}\sqrt{\beta}}e^{i\omega_{n}\tau+i\vec{k}\cdot\vec{x}}
\end{equation}
where $\omega_{n}=2\pi n/\beta$ for $n=0,\pm 1,\pm 2,\dots$, and
\begin{equation}
\int_{0}^{\beta}d\tau\int d^{N-1}x\ \! \phi^{*}_{n'\vec{k}'}(\tau,\vec{x})\phi_{n\vec{k}}(\tau,\vec{x})=\delta_{nn'}\ \!\delta(\vec{k}-\vec{k}')
\end{equation}

From the tensorial structure of the $\Delta T_{\mu\nu\alpha'\beta'}^{2}(x,x')$, one can define the scalar coefficient functions as follows \cite{AllJac86,PRVdS,ChoHuAdS}.
\begin{eqnarray}
\Delta T_{0000}(x,x')&=&C_{11}\label{cor0000}\\
\Delta T_{000i}(x,x')&=&C_{21}(\vec{x}-\vec{x})_{i}\\
\Delta T_{00ij}(x,x')&=&C_{31}\delta_{ij}+C_{32}(\vec{x}-\vec{x}')_{i}(\vec{x}-\vec{x}')_{j}\\
\Delta T_{0i0j}(x,x')&=&C_{41}\delta_{ij}+C_{42}(\vec{x}-\vec{x}')_{i}(\vec{x}-\vec{x}')_{j}\\
\Delta T_{0ijk}(x,x')&=&C_{51}(\vec{x}-\vec{x}')_{i}\delta_{jk}+C_{52}
\left[\delta_{ij}(\vec{x}-\vec{x}')_{k}+\delta_{ik}(\vec{x}-\vec{x}')_{j}\right]\nonumber\\
&&\ \ +C_{53}(\vec{x}-\vec{x}')_{i}(\vec{x}-\vec{x}')_{j}(\vec{x}-\vec{x}')_{k}
\end{eqnarray}
\begin{eqnarray}
\Delta T_{ijkl}(x,x')&=&C_{61}\delta_{ij}\delta_{kl}+C_{62}(\delta_{ik}\delta_{jl}+\delta_{il}\delta_{jk})\nonumber\\
&&\ \ +C_{63}\left[\delta_{ij}(\vec{x}-\vec{x}')_{k}(\vec{x}-\vec{x}')_{l}+\delta_{kl}(\vec{x}-\vec{x}')_{i}(\vec{x}-\vec{x}')_{j}\right]\nonumber\\
&&\ \ +C_{64}\left[\delta_{ik}(\vec{x}-\vec{x}')_{j}(\vec{x}-\vec{x}')_{l}+\delta_{il}(\vec{x}-\vec{x}')_{j}(\vec{x}-\vec{x}')_{k}\right.\nonumber\\
&&\ \ \ \ \ \ \ \ \ \ \ \left.+\delta_{jk}(\vec{x}-\vec{x}')_{i}(\vec{x}-\vec{x}')_{l}+\delta_{jl}(\vec{x}-\vec{x}')_{i}(\vec{x}-\vec{x}')_{k}\right]\nonumber\\
&&\ \ +C_{65}(\vec{x}-\vec{x}')_{i}(\vec{x}-\vec{x}')_{j}
(\vec{x}-\vec{x}')_{k}(\vec{x}-\vec{x}')_{l}\label{corijkl}
\end{eqnarray}
where, owing to the homogeneity of the space, the coefficients $C_{ab}$ are functions of $\tau-\tau'$ and $|\vec{x}-\vec{x}'|$ only.

The components of $T_{\mu\nu}[\phi_{n\vec{k}}(\tau,\vec{x}),\phi^{*}_{n'\vec{k}'}(\tau',\vec{x}')]$ in hot flat space are
\begin{eqnarray}
T_{00}[\phi_{n\vec{k}}(\tau,\vec{x}),\phi^{*}_{n'\vec{k}'}(\tau',\vec{x}')]&=&\left[-\omega_{n}\left(\omega_{n}+\omega_{n'}\right)
-\vec{k}\cdot(\vec{k}-\vec{k}')\right]\phi_{n\vec{k}}(\tau,\vec{x}),\phi^{*}_{n'\vec{k}'}(\tau',\vec{x}')\\
T_{0i}[\phi_{n\vec{k}}(\tau,\vec{x}),\phi^{*}_{n'\vec{k}'}(\tau',\vec{x}')]&=&\left[-\omega_{n}k_{i}'-\omega_{n'}k_{i}\right]
\phi_{n\vec{k}}(\tau,\vec{x}),\phi^{*}_{n'\vec{k}'}(\tau',\vec{x}')\\
T_{ij}[\phi_{n\vec{k}}(\tau,\vec{x}),\phi^{*}_{n'\vec{k}'}(\tau',\vec{x}')]&=&\left[-k_{i}k_{j}'-k_{i}'k_{j}-\delta_{ij}
\left(\omega_{n}\left(\omega_{n}-\omega_{n'}\right)+\vec{k}\cdot(\vec{k}-\vec{k}')\right)\right]\times\nonumber\\
&&\ \ \phi_{n\vec{k}}(\tau,\vec{x}),\phi^{*}_{n'\vec{k}'}(\tau',\vec{x}')
\end{eqnarray}
The coefficients $C_{ab}$ can be expressed in terms of the function $f(\alpha)$,
\begin{equation}
f(\alpha)=\sum_{n=-\infty}^{\infty}\int_{0}^{\infty}du\ \! u^{-\alpha}e^{-u\omega_{n}^{2}}e^{-(\vec{x}-\vec{x}')^{2}/4u}e^{i\omega_{n}(\tau-\tau')}
\label{fdef}
\end{equation}
and its $\tau$ derivatives. These expressions can be found in Appendix \ref{appa}.

\section{Low and high temperature expansions of the correlators}

To develop the low and high temperature expansions of the correlators, we look at the corresponding behavior of the function $f(\alpha)$ defined in Eq.~(\ref{fdef}). In fact, for the low temperature regime, a Poisson summation formula has to be used to obtain the appropriate form of $f(\alpha)$. In these expansions we shall see that the dependence of the correlators on the temperature are quite different in the low and the high temperature regimes. We shall also discuss the corresponding singular behaviors of the correlators in the $x\rightarrow x'$ coincident limit.

\subsection{High temperature expansion}

First for high temperature or small $\beta$, $\omega_{n}=2\pi n/\beta$ is always large except for $n=0$. Hence, we need to treat the $n=0$ term separately, as follows:
\begin{eqnarray}
f(\alpha)
&=&\int_{0}^{\infty}du\ \! u^{-\alpha}e^{-(\vec{x}-\vec{x}')^{2}/4u}+2\sum_{n=1}^{\infty}\cos[\omega_{n}(\tau-\tau')]
\int_{0}^{\infty}du\ \! u^{-\alpha}e^{u\omega_{n}^{2}}e^{(\vec{x}-\vec{x}')^{2}/4u}\nonumber\\
&=&4^{\alpha-1}\Gamma(\alpha-1)[(\vec{x}-\vec{x}')^{2}]^{1-\alpha}\nonumber\\
&&+\sum_{n=1}^{\infty}2^{\alpha+1}[(\vec{x}-\vec{x}')^{2}]^{1-\alpha}\left(\frac{2\pi n|\vec{x}-\vec{x}'|}{\beta}\right)^{\alpha-1}
\cos\left(\frac{2\pi n(\tau-\tau')}{\beta}\right)K_{\alpha-1}\left(\frac{2\pi n|\vec{x}-\vec{x}'|}{\beta}\right)\nonumber\\
\end{eqnarray}
Using the asymptotic expansion of the modified Bessel function $K_{\nu}(z)$ for large $z$, we can further expand $f(\alpha)$.
\begin{eqnarray}
f(\alpha)&=&4^{\alpha-1}\Gamma(\alpha-1)[(\vec{x}-\vec{x}')^{2}]^{1-\alpha}\times\nonumber\\
&&\left\{
1+\frac{2\pi^{\alpha-1}}{\Gamma(\alpha-1)}e^{-2\pi|\vec{x}-\vec{x}'|/\beta}\left(\frac{|\vec{x}-\vec{x}'|}{\beta}\right)^{\alpha-\frac{3}{2}}
\cos\left(\frac{2\pi(\tau-\tau')}{\beta}\right)\right.\nonumber\\
&&\ \ \ \ \ \ \ \ \ \left.\left[1+\frac{\left(\alpha-\frac{1}{2}\right)\left(\alpha-\frac{3}{2}\right)}{4\pi}
\left(\frac{\beta}{|\vec{x}-\vec{x}'|}\right)+\cdots\right]+\cdots\right\}\label{fhight}
\end{eqnarray}
This asymptotic series of $f(\alpha)$ can now be used to develop the high temperature expansions of the correlator components. They are listed in Appendix~\ref{appb}. From Eqs.~(\ref{high1}) to (\ref{high6}), we see that  the density-density correlator $\Delta T_{0000}^{2}(x,x')$, the density-pressure correlators $\Delta T_{00ij}^{2}(x,x')$ and the pressure-pressure correlators $\Delta T_{ijkl}^{2}(x,x')$  are of the order of $1/\beta^{2}$ or $T^{2}$, while $\Delta T_{000i}^{2}(x,x')$, $\Delta T_{0i0j}^{2}(x,x')$ and $\Delta T_{0ijk}^{2}(x,x')$ are suppressed by $e^{-T}$. 

Next, we investigate the singular behaviors of the correlators in the $x'\rightarrow x$ coincident limit. Since the space is homogeneous and isotropic, we need to average over the directions in this limit. This amounts to replacing $(\vec{x}-\vec{x}')_{i}(\vec{x}-\vec{x}')_{j}$ with $\delta_{ij}(\vec{x}-\vec{x}')^{2}/(N-1)$ and $(\vec{x}-\vec{x}')_{i}(\vec{x}-\vec{x}')_{j}(\vec{x}-\vec{x}')_{k}(\vec{x}-\vec{x}')_{l}$ with $(\delta_{ij}\delta_{kl}+\delta_{ik}\delta_{jl}+\delta_{il}\delta_{jk})(\vec{x}-\vec{x}')^{4}/(N-1)^{2}$. In this way, the correlators are
\begin{eqnarray}
\Delta T_{0000}^{2}(x,x')&\sim&\frac{(N-1)(N-2)\Gamma^{2}[(N-1)/2]}{8\pi^{N-1}|\vec{x}-\vec{x}'|^{2N-2}\beta^{2}}\\
\Delta T_{00ij}^{2}(x,x')&\sim&\frac{(N-2)(N-3)\Gamma^{2}[(N-1)/2]\delta_{ij}}{8\pi^{N-1}|\vec{x}-\vec{x}'|^{2N-2}\beta^{2}}\\
\Delta T_{ijkl}^{2}(x,x')&\sim&\frac{(N^{2}-7N+14)\Gamma^{2}[(N-1)/2]}{8\pi^{N-1}|\vec{x}-\vec{x}'|^{2N-2}\beta^{2}}
\left[\delta_{ij}\delta_{kl}+\frac{2}{N^{2}-7N+14}\left(\delta_{ik}\delta_{jl}+\delta_{il}\delta_{jk}\right)\right]\nonumber\\
\end{eqnarray}
For example, for $N=4$, we have
\begin{eqnarray}
\Delta T_{0000}^{2}(x,x')&\sim&\frac{3}{16\pi^{2}}\left(\frac{1}{|\vec{x}-\vec{x}'|^{6}\beta^{2}}\right)\\
\Delta T_{00ij}^{2}(x,x')&\sim&\frac{1}{16\pi^{2}}\left(\frac{1}{|\vec{x}-\vec{x}'|^{6}\beta^{2}}\right)\delta_{ij}\\
\Delta T_{ijkl}^{2}(x,x')&\sim&\frac{1}{16\pi^{2}}\left(\frac{1}{|\vec{x}-\vec{x}'|^{6}\beta^{2}}\right)
\left(\delta_{ij}\delta_{kl}+\delta_{ik}\delta_{jl}+\delta_{il}\delta_{jk}\right)
\end{eqnarray}
The density-density fluctuation is much larger than the other fluctuations. However, when $N$ is large, all these three fluctuations have the same $N^{2}\Gamma^{2}[(N-1)/2]$ dependence.

\subsection{Low temperature expansion}

To find the low temperature expansion for the correlators we again consider the function $f(\alpha)$. The definition in Eq.~(\ref{fdef}) is not suitable for this purpose. An appropriate form can be obtained using the Poisson summation formula (See, for example, \cite{Camp90}),
\begin{equation}
\sum_{n=-\infty}^{\infty}g(n\beta)=\sum_{n=-\infty}^{\infty}\frac{\sqrt{2\pi}}{\beta}\tilde{g}\left(\frac{2\pi n}{\beta}\right)
\end{equation}
where $\tilde{g}(k)$ is the Fourier transform of $g(x)$,
\begin{equation}
\tilde{g}(k)=\int_{-\infty}^{\infty}\frac{dx}{\sqrt{2\pi}}e^{-ikx}g(x)
\end{equation}
Taking $g(x)=e^{-4\pi^{2}ux^{2}/\beta^{2}+i2\pi(\tau-\tau')x/\beta^{2}}$, we have
\begin{equation}
\sum_{n=-\infty}^{\infty}e^{-u\omega_{n}^{2}+i\omega_{n}(\tau-\tau')}
=\sum_{n=-\infty}^{\infty}\frac{\beta}{\sqrt{4\pi u}}e^{(\tau-\tau'-n\beta)^{2}/4u}
\end{equation}
and the function $f(\alpha)$ becomes
\begin{equation}
f(\alpha)=\sum_{n=-\infty}^{\infty}\frac{2^{2\alpha-1}\Gamma\left(\alpha-\frac{1}{2}\right)}{\sqrt{4\pi}}
\left[(\vec{x}-\vec{x}')^{2}+(\tau-\tau'-n\beta)^{2}\right]^{-\alpha+\frac{1}{2}}
\end{equation}
In this form one can expand in powers of $1/\beta$ for large $\beta$ or low temperature.
\begin{eqnarray}
f(\alpha)&=&\frac{2^{2\alpha-1}\Gamma\left(\alpha-\frac{1}{2}\right)}{\sqrt{4\pi}}
\left[(\vec{x}-\vec{x}')^{2}+(\tau-\tau')^{2}\right]^{-\alpha+\frac{1}{2}}
+\frac{2^{2\alpha-1}\Gamma\left(\alpha-\frac{1}{2}\right)\zeta\left(2\alpha-1\right)}{\sqrt{\pi}\beta^{2\alpha-2}}\times\nonumber\\
&&\ \ \ \ \ \ \ \ \ \ \left\{1-\frac{(2\alpha-1)\zeta\left(2\alpha+1\right)}{2\zeta(2\alpha-1)\beta^{2}}
\left[(\vec{x}-\vec{x}')^{2}-2\alpha(\tau-\tau')^{2}\right]+\cdots\right\}\nonumber\\\label{flowt}
\end{eqnarray}
Using this expansion we can develop the low temperature expansions of the correlators. They are listed in Appendix~\ref{appc}. We make two observations: 1) From Eqs.~(\ref{low1}) to (\ref{low6}), we see that when $\beta\rightarrow\infty$ or $T\rightarrow 0$, the correlators reduce to the ones in a $R^{N}$ space. One can check that the values we obtain here are the same as the ones in \cite{ChoHuAdS}. 2) The leading temperature dependence of the correlators are all of $1/\beta^{N}$ or $T^{N}$.

To investigate the singular behaviors of the correlators in the coincident limit in this low temperature regime, we first take the $\tau'\rightarrow\tau$ limit. In this limit $\Delta T_{000i}^{2}$ and $\Delta T_{0ijk}^{2}$ vanish. We then average over the directions in the other correlators giving
\begin{eqnarray}
\Delta T_{0000}^{2}(x,x')&\sim&\frac{N(N-1)\Gamma^{2}(N/2)}{8\pi^{N}|\vec{x}-\vec{x}'|^{2N}}
\left[1-\frac{4\zeta(N)}{N-1}\left(\frac{|\vec{x}-\vec{x}'|}{\beta}\right)^{N}\right]\\
\Delta T_{00ij}^{2}(x,x')&\sim&\frac{(N^{3}-4N^{2}+N+4)\Gamma^{2}(N/2)\delta_{ij}}{8(N-1)\pi^{N}|\vec{x}-\vec{x}'|^{2N}}
\left[1+\frac{4(N^{2}-3N+4)\zeta(N)}{(N^{3}-4N^{2}+N+4)}\left(\frac{|\vec{x}-\vec{x}'|}{\beta}\right)^{N}\right]\nonumber\\ \\
\Delta T_{0i0j}^{2}(x,x')&\sim&-\frac{\Gamma^{2}(N/2)\delta_{ij}}{4(N-1)\pi^{N}|\vec{x}-\vec{x}'|^{2N}}
\left[1-4(N-1)\zeta(N)\left(\frac{|\vec{x}-\vec{x}'|}{\beta}\right)^{N}\right]\\
\Delta T_{ijkl}^{2}(x,x')&\sim&\frac{(N^{4}-7N^{3}+15N^{2}-N-4)\Gamma^{2}(N/2)(\delta_{ij}\delta_{kl})}{8(N-1)^{2}\pi^{N}|\vec{x}-\vec{x}'|^{2N}}
\times\nonumber\\
&&\ \ \ \ \ \left[1-\frac{4(N-1)(N^{2}+7N-4)\zeta(N)}{(N^{4}-7N^{3}+15N^{2}-N-4)}\left(\frac{|\vec{x}-\vec{x}'|}{\beta}\right)^{N}\right]\\
&&+\frac{(N^{2}+1)\Gamma^{2}(N/2)(\delta_{ik}\delta_{jl}+\delta_{il}\delta_{jk})}{4(N-1)^{2}\pi^{N}|\vec{x}-\vec{x}'|^{2N}}
\left[1-\frac{4(N-1)\zeta(N)}{(N^{2}+1)}\left(\frac{|\vec{x}-\vec{x}'|}{\beta}\right)^{N}\right]
\end{eqnarray}
From the signs of the temperature dependent terms we see that the finite temperature effect tends to increase the density-pressure fluctuation while it tends to decrease all the other fluctuations.

To see their relative magnitudes we examine the four dimensional case. Setting $N=4$ in the above we get
\begin{eqnarray}
\Delta T_{0000}^{2}(x,x')&\sim&\frac{3}{2\pi^{4}|\vec{x}-\vec{x}'|^{8}}
\left[1-\frac{2\pi^{4}}{135}\left(\frac{|\vec{x}-\vec{x}'|}{\beta}\right)^{4}\right]\\
\Delta T_{00ij}^{2}(x,x')&\sim&\frac{\delta_{ij}}{3\pi^{4}|\vec{x}-\vec{x}'|^{8}}
\left[1+\frac{2\pi^{4}}{45}\left(\frac{|\vec{x}-\vec{x}'|}{\beta}\right)^{4}\right]
\end{eqnarray}
\begin{eqnarray}
\Delta T_{0i0j}^{2}(x,x')&\sim&-\frac{\delta_{ij}}{12\pi^{4}|\vec{x}-\vec{x}'|^{8}}
\left[1-\frac{2\pi^{4}}{15}\left(\frac{|\vec{x}-\vec{x}'|}{\beta}\right)^{4}\right]\\
\Delta T_{ijkl}^{2}(x,x')&\sim&\frac{5\delta_{ij}\delta_{kl}}{9\pi^{4}|\vec{x}-\vec{x}'|^{8}}
\left[1-\frac{2\pi^{4}}{15}\left(\frac{|\vec{x}-\vec{x}'|}{\beta}\right)^{4}\right]\nonumber\\
&&\ \ \ \ \ +
\frac{17(\delta_{ik}\delta_{jl}+\delta_{il}\delta_{jk})}{36\pi^{4}|\vec{x}-\vec{x}'|^{8}}
\left[1-\frac{2\pi^{4}}{255}\left(\frac{|\vec{x}-\vec{x}'|}{\beta}\right)^{4}\right]
\end{eqnarray}
Here the finite temperature effect is larger for the current-current and the pressure-pressure fluctuations than the others. On the other hand, for large values $N$, the current-current fluctuation goes like $4N\zeta(N)$, while the other fluctuations all go like $4\zeta(N)/N$.

\section{Concluding Remarks}

In this paper we have calculated the stress-energy tensor correlators of a massless, minimally coupled scalar quantum field in hot N-dimensional flat spaces. With the help of the Poisson summation formula we are able to develop low temperature as well as high temperature expansions of these correlators. Low and high temperature regimes are determined by whether $T$ is smaller or larger than $|\vec{x}-\vec{x}'|$, the only dimensional parameter in the problem. From these expansions we see that the correlators have rather different behaviors in these regimes. For low temperature the correlators all have finite temperature corrections of the order $T^{N}$. However, for high temperature, the density-density, the density-pressure and the pressure-pressure correlators are of the order $T^{2}$, while the other correlators are suppressed by $e^{-T}$.

We have also investigated the quantum fluctuations of the stress-energy tensor by considering the singular behaviors of the correlators under the $x'\rightarrow x$ coincident limit. Here we first take $\tau'\rightarrow\tau$ and then average over the directions. In the low temperature regime we see that the finite temperature contributions tend to decrease the magnitude of fluctuation because the vacuum and the thermal fluctuations are of opposite signs, except for $\Delta T_{00ij}^{2}$. On the other hand, in the high temperature regime, the unsuppressed correlators all have similar magnitude which go like $N^{2}\Gamma^{2}[(N-1)/2]$ for large $N$. Note that in this limit, both $\beta=1/T$ and $|\vec{x}-\vec{x}'|$ are small but $|\vec{x}-\vec{x}'|>\beta$ is always assumed in our approximation.

It would be interesting to see how these behaviors would change with the introduction of another dimensional parameter. For example, we can consider massive scalars or background spacetime with non-zero curvature, like in \cite{ChoHuAdS}. We do not expect the high temperature behavior to change. However, the low temperature behavior might be altered. This makes careful consideration of finite yet small temperature corrections to established results involving quantum vacuum quantities, such as negative energy density and theorems derived therefrom, such as the quantum interest principle, worth its while. We hope to address these foundational issues of importance and report our findings certainly on, if not before, the occasion of Professor Dowker's 80th birthday celebration.

Here's to wishing you, Stuart, many happy returns!

\acknowledgments  HTC is supported in part by the National Science Council of the Republic of China under the Grant NSC 99-2112-M-032-003-MY3, and by the National Center for Theoretical Sciences. BLH is supported in part by the US National Science Foundation under grant PHY-0801368 to the University of Maryland. This paper was finished while he was a visiting professor at the Changhua University of Education, Taiwan and the University of Science and Technology, Hong Kong.

\appendix

\section{Coefficient functions}\label{appa}

The coefficient functions of the stress-energy correlators in Eqs.~(\ref{cor0000}) to (\ref{corijkl}):
\begin{eqnarray}
&&C_{11}(\tau-\tau',|\vec{x}-\vec{x}'|)\nonumber\\
&=&\frac{1}{2(4\pi)^{N-1}\beta^{2}}\times\nonumber\\
&&\left\{2\partial_{\tau}^{2}f(\frac{N-1}{2})\partial_{\tau}^{2}f(\frac{N-1}{2})
-2\partial_{\tau}^{3}f(\frac{N-1}{2})\partial_{\tau}f(\frac{N-1}{2})\right.\nonumber\\
&&\ \ +(N-1)\left[\partial_{\tau}f(\frac{N-1}{2})\partial_{\tau}f(\frac{N-1}{2}+1)
-f(\frac{N-1}{2}+1)\partial_{\tau}^{2}f(\frac{N-1}{2})\right]\nonumber\\
&&\ \ \ \ \ +\frac{N(N-1)}{4}f(\frac{N-1}{2}+1)f(\frac{N-1}{2}+1)\nonumber\\
&&\ \ -\frac{(\vec{x}-\vec{x}')^{2}}{2}\left[\partial_{\tau}f(\frac{N-1}{2})\partial_{\tau}f(\frac{N-1}{2}+2)
-f(\frac{N-1}{2}+2)\partial_{\tau}^{2}f(\frac{N-1}{2})\right.\nonumber\\
&&\ \ \ \ \ \ \ \ \partial_{\tau}f(\frac{N-1}{2}+1)\partial_{\tau}f(\frac{N-1}{2}+1)
-f(\frac{N-1}{2}+1)\partial_{\tau}^{2}f(\frac{N-1}{2}+1)\nonumber\\
&&\ \ \ \ \ \ \ \ \ \left.+\frac{2N+1}{2}f(\frac{N-1}{2}+1)f(\frac{N-1}{2}+2)\right]\nonumber\\
&&\ \ \left.+\frac{(\vec{x}-\vec{x}')^{4}}{8}\left[f(\frac{N-1}{2}+2)f(\frac{N-1}{2}+2)+
f(\frac{N-1}{2}+1)f(\frac{N-1}{2}+3)\right]\right\}\nonumber\\
\end{eqnarray}
\begin{eqnarray}
&&C_{21}(\tau-\tau',|\vec{x}-\vec{x'}|)\nonumber\\
&=&\frac{1}{2(4\pi)^{N-1}\beta^{2}}\times\nonumber\\
&&\left\{\frac{1}{2}\partial_{\tau}^{2}f(\frac{N-1}{2}+1)\partial_{\tau}f(\frac{N-1}{2})
-\frac{1}{2}\partial_{\tau}f(\frac{N-1}{2}+1)\partial_{\tau}^{2}f(\frac{N-1}{2})\right.\nonumber\\
&&\ \ +\frac{1}{2}f(\frac{N-1}{2}+1)\partial_{\tau}^{3}f(\frac{N-1}{2})
-\frac{1}{2}\partial_{\tau}^{2}f(\frac{N-1}{2})\partial_{\tau}f(\frac{N-1}{2}+1)\nonumber\\
&&\ \ -\frac{N+1}{4}\left[f(\frac{N-1}{2}+1)\partial_{\tau}f(\frac{N-1}{2}+1)
+f(\frac{N-1}{2}+2)\partial_{\tau}f(\frac{N-1}{2})\right]\nonumber\\
&&+\frac{(\vec{x}-\vec{x}')^{2}}{4}\left[f(\frac{N-1}{2}+2)\partial_{\tau}f(\frac{N-1}{2}+1)
\right.\nonumber\\
&&\ \ \left.\left.+\frac{1}{2}f(\frac{N-1}{2}+1)\partial_{\tau}f(\frac{N-1}{2}+2)
+\frac{1}{2}f(\frac{N-1}{2}+3)\partial_{\tau}f(\frac{N-1}{2})\right]\right\}\nonumber\\
\end{eqnarray}
\begin{eqnarray}
&&C_{31}(\tau-\tau',|\vec{x}-\vec{x}'|)\nonumber\\
&=&\frac{1}{2(4\pi)^{N-1}\beta^{2}}\times\nonumber\\
&&\left\{\frac{N^{2}-N-2}{4}f(\frac{N-1}{2}+1)f(\frac{N-1}{2}+1)
-(N-1)f(\frac{N-1}{2}+1)\partial_{\tau}^{2}f(\frac{N-1}{2})\right.\nonumber\\
&&+\frac{(\vec{x}-\vec{x}')^2}{4}\left[2f(\frac{N-1}{2}+2)\partial_{\tau}^{2}f(\frac{N-1}{2})
+2f(\frac{N-1}{2}+1)\partial_{\tau}^{2}f(\frac{N-1}{2}+1)\right.\nonumber\\
&&\ \ \left.-(2N+1)f(\frac{N-1}{2}+2)f(\frac{N-1}{2}+1)\right]\nonumber\\
&&\left.+\frac{(\vec{x}-\vec{x}')^{4}}{8}\left[f(\frac{N-1}{2}+3)f(\frac{N-1}{2}+1)
+f(\frac{N-1}{2}+2)f(\frac{N-1}{2}+2)\right]\right\}\nonumber\\
\end{eqnarray}
\begin{eqnarray}
&&C_{32}(\tau-\tau',|\vec{x}-\vec{x}'|)\nonumber\\
&=&\frac{1}{2(4\pi)^{N-1}\beta^{2}}\times\nonumber\\
&&\left\{\frac{1}{2}\partial_{\tau}f(\frac{N-1}{2}+1)\partial_{\tau}f(\frac{N-1}{2}+1)
-\frac{1}{2}f(\frac{N-1}{2}+1)\partial_{\tau}^{2}f(\frac{N-1}{2}+1)\right.\nonumber\\
&&\ \ +\frac{N+3}{4}f(\frac{N-1}{2}+2)f(\frac{N-1}{2}+1)\nonumber\\
&&\left.-\frac{(\vec{x}-\vec{x}')^{2}}{8}\left[f(\frac{N-1}{2}+3)f(\frac{N-1}{2}+1)
+f(\frac{N-1}{2}+2)f(\frac{N-1}{2}+2)\right]\right\}\nonumber\\
\end{eqnarray}
\begin{eqnarray}
&&C_{41}(\tau-\tau',|\vec{x}-\vec{x}'|)\nonumber\\
&=&\frac{1}{2(4\pi)^{N-1}\beta^{2}}\left\{-f(\frac{N-1}{2}+1)\partial_{\tau}^{2}f(\frac{N-1}{2})\right\}\nonumber\\
\\
&&C_{42}(\tau-\tau',|\vec{x}-\vec{x}'|)\nonumber\\
&=&\frac{1}{2(4\pi)^{N-1}\beta^{2}}\times\nonumber\\
&&\left\{\frac{1}{2}\partial_{\tau}f(\frac{N-1}{2}+1)\partial_{\tau}f(\frac{N-1}{2}+1)
+\frac{1}{2}f(\frac{N-1}{2}+2)\partial_{\tau}^{2}f(\frac{N-1}{2})\right\}\nonumber\\
\end{eqnarray}
\begin{eqnarray}
&&C_{51}(\tau-\tau',|\vec{x}-\vec{x}'|)\nonumber\\
&=&\frac{1}{2(4\pi)^{N-1}\beta^{2}}\times\nonumber\\
&&\left\{\partial_{\tau}f(\frac{N-1}{2}+1)\partial_{\tau}^{2}f(\frac{N-1}{2})
+\frac{1}{2}\partial_{\tau}f(\frac{N-1}{2})\partial_{\tau}^{2}f(\frac{N-1}{2}+1)\right.\nonumber\\
&&\ \ +\frac{1}{2}f(\frac{N-1}{2}+1)\partial_{\tau}^{3}f(\frac{N-1}{2})
-\frac{N+1}{4}f(\frac{N-1}{2}+2)\partial_{\tau}f(\frac{N-1}{2})\nonumber\\
&&\ \ -\frac{N+1}{4}f(\frac{N-1}{2}+1)\partial_{\tau}f(\frac{N-1}{2}+1)\nonumber\\
&&+\frac{(\vec{x}-\vec{x}')^{2}}{4}\left[f(\frac{N-1}{2}+2)\partial_{\tau}f(\frac{N-1}{2}+1)
+\frac{1}{2}f(\frac{N-1}{2}+3)\partial_{\tau}f(\frac{N-1}{2})\right.\nonumber\\
&&\ \ \left.\left.+\frac{1}{2}f(\frac{N-1}{2}+1)\partial_{\tau}f(\frac{N-1}{2}+2)\right]\right\}\nonumber\\ \\
&&C_{52}(\tau-\tau',|\vec{x}-\vec{x}'|)\nonumber\\
&=&\frac{1}{2(4\pi)^{N-1}\beta^{2}}\left\{\frac{1}{2}f(\frac{N-1}{2}+1)\partial_{\tau}f(\frac{N-1}{2}+1)\right\}\nonumber\\ \\
&&C_{53}(\tau-\tau',|\vec{x}-\vec{x}'|)\nonumber\\
&=&\frac{1}{2(4\pi)^{N-1}\beta^{2}}\left\{-\frac{1}{2}f(\frac{N-1}{2}+2)\partial_{\tau}f(\frac{N-1}{2}+1)\right\}
\end{eqnarray}
\begin{eqnarray}
&&C_{61}(\tau-\tau',|\vec{x}-\vec{x}'|)\nonumber\\
&=&\frac{1}{2(4\pi)^{N-1}\beta^{2}}\times\nonumber\\
&&\left\{2\partial_{\tau}f(\frac{N-1}{2})\partial_{\tau}^{3}f(\frac{N-1}{2})
+2\partial_{\tau}^{2}f(\frac{N-1}{2})\partial_{\tau}^{2}f(\frac{N-1}{2})\right.\nonumber\\
&&-(N-1)f(\frac{N-1}{2}+1)\partial_{\tau}^{2}f(\frac{N-1}{2})
-(N-1)\partial_{\tau}f(\frac{N-1}{2})\partial_{\tau}f(\frac{N-1}{2}+1)\nonumber\\
&&+\frac{N^{2}-N-4}{4}f(\frac{N-1}{2}+1)f(\frac{N-1}{2}+1)\nonumber\\
&&+\frac{(\vec{x}-\vec{x}')^{2}}{2}\left[f(\frac{N-1}{2}+1)\partial_{\tau}^{2}f(\frac{N-1}{2}+1)
+\partial_{\tau}f(\frac{N-1}{2}+1)\partial_{\tau}f(\frac{N-1}{2}+1)\right.\nonumber\\
&&\ \ +f(\frac{N-1}{2}+2)\partial_{\tau}^{2}f(\frac{N-1}{2})
+\partial_{\tau}f(\frac{N-1}{2})\partial_{\tau}f(\frac{N-1}{2}+2)\nonumber\\
&&\ \ \left.-\frac{2N+1}{2}f(\frac{N-1}{2}+2)f(\frac{N-1}{2}+1)\right]\nonumber\\
&&\left.+\frac{(\vec{x}-\vec{x}')^{4}}{8}\left[f(\frac{N-1}{2}+2)f(\frac{N-1}{2}+2)
+f(\frac{N-1}{2}+3)f(\frac{N-1}{2}+1)\right]\right\}\nonumber\\
\end{eqnarray}
\begin{eqnarray}
&&C_{62}(\tau-\tau',|\vec{x}-\vec{x}'|)\nonumber\\
&=&\frac{1}{2(4\pi)^{N-1}\beta^{2}}\left\{\frac{1}{2}f(\frac{N-1}{2}+1)f(\frac{N-1}{2}+1)\right\}\nonumber\\ \\
&&C_{63}(\tau-\tau',|\vec{x}-\vec{x}'|)\nonumber\\
&=&\frac{1}{2(4\pi)^{N-1}\beta^{2}}\times\nonumber\\
&&\left\{-\frac{1}{2}f(\frac{N-1}{2}+1)\partial_{\tau}^{2}f(\frac{N-1}{2}+1)
-\frac{1}{2}\partial_{\tau}f(\frac{N-1}{2}+1)\partial_{\tau}f(\frac{N-1}{2}+1)\right.\nonumber\\
&&\ \ \frac{N+3}{4}f(\frac{N-1}{2}+2)f(\frac{N-1}{2}+1)\nonumber\\
&&\left.-\frac{(\vec{x}-\vec{x}')^{2}}{8}\left[f(\frac{N-1}{2}+3)f(\frac{N-1}{2}+1)
+f(\frac{N-1}{2}+2)f(\frac{N-1}{2}+2)\right]\right\}\nonumber\\
\\
&&C_{64}(\tau-\tau',|\vec{x}-\vec{x}'|)\nonumber\\
&=&\frac{1}{2(4\pi)^{N-1}\beta^{2}}\left\{-\frac{1}{4}f(\frac{N-1}{2}+2)f(\frac{N-1}{2}+1)\right\}\nonumber\\
\end{eqnarray}
\begin{eqnarray}
&&C_{65}(\tau-\tau',|\vec{x}-\vec{x}'|)\nonumber\\
&=&\frac{1}{2(4\pi)^{N-1}\beta^{2}}\left\{\frac{1}{4}f(\frac{N-1}{2}+2)f(\frac{N-1}{2}+2)\right\}\nonumber\\
\end{eqnarray}

\section{High temperature expansion}\label{appb}

Using Eq.~(\ref{fhight}) we have the high temperature expansions of the correlators
\begin{eqnarray}
&&\Delta T_{0000}^{2}(x,x')\nonumber\\
&=&\frac{(N-1)(N-2)\Gamma^{2}[(N-1)/2]}{8\pi^{N-1}|\vec{x}-\vec{x}'|^{2N}}\left(\frac{|\vec{x}-\vec{x}'|}{\beta}\right)^{2}\nonumber\\
&&+e^{-2\pi|\vec{x}-\vec{x}'|/\beta}\cos\left(\frac{2\pi(\tau-\tau')}{\beta}\right)\frac{(N-2)\Gamma[(N-1)/2]}{\pi^{(N-3)/2}|\vec{x}-\vec{x}'|^{2N}}
\left(\frac{|\vec{x}-\vec{x'}|}{\beta}\right)^{\frac{N}{2}+2}\times\nonumber\\
&&\ \ \ \ \ \left[1-\frac{3N(N+2)}{16(N-2)\pi}\left(\frac{\beta}{|\vec{x}-\vec{x}'|}\right)+\cdots\right]+\cdots\nonumber\\ \label{high1}\\
&&\Delta T_{000i}^{2}(x,x')\nonumber\\
&=&-\frac{(\vec{x}-\vec{x}')_{i}}{\beta}e^{-2\pi|\vec{x}-\vec{x}'|/\beta}\sin\left(\frac{2\pi(\tau-\tau')}{\beta}\right)
\frac{(N-2)\Gamma[(N-1)/2]}{\pi^{(N-3)/2}|\vec{x}-\vec{x}'|^{2N}}\times\nonumber\\
&&\ \ \ \ \ \left(\frac{|\vec{x}-\vec{x'}|}{\beta}\right)^{\frac{N}{2}+1}
\left[1-\frac{N(N-3)}{32\pi}\left(\frac{\beta}{|\vec{x}-\vec{x}'|}\right)+\cdots\right]+\cdots\nonumber\\  \\
&&\Delta T_{00ij}^{2}(x,x')\nonumber\\
&=&\left[N\frac{|\vec{x}-\vec{x}'|^{2}}{\beta^{2}}\delta_{ij}
-2(N-1)\frac{(\vec{x}-\vec{x}')_{i}(\vec{x}-\vec{x}')_{j}}{\beta^{2}}\right]
\frac{(N-3)\Gamma^{2}[(N-1)/2]}{8\pi^{N-1}|\vec{x}-\vec{x}'|^{2N}}\nonumber\\
&&+e^{-2\pi|\vec{x}-\vec{x}'|/\beta}\cos\left(\frac{2\pi(\tau-\tau')}{\beta}\right)\frac{(N-2)\Gamma[(N-1)/2]}{\pi^{(N-3)/2}|\vec{x}-\vec{x}'|^{2N}}
\left(\frac{|\vec{x}-\vec{x'}|}{\beta}\right)^{N/2}\times\nonumber\\
&&\ \ \left\{\delta_{ij}\left(\frac{|\vec{x}-\vec{x}'|^{2}}{\beta^{2}}\right)
\left[1-\frac{3N^{2}+6N+16}{16(N-2)\pi}\left(\frac{\beta}{|\vec{x}-\vec{x}'|}\right)+\cdots\right]\right.\nonumber\\
&&\ \ \ \ \ \left.-2\frac{(\vec{x}-\vec{x}')_{i}(\vec{x}-\vec{x}')_{j}}{\beta^{2}}
\left[1+\frac{N^{3}-3N^{2}-26N+16}{32(N-2)\pi}\left(\frac{\beta}{|\vec{x}-\vec{x}'|}\right)+\cdots\right]\right\}\nonumber\\
&&+\cdots\nonumber\\
\end{eqnarray}
\begin{eqnarray}
&&\Delta T_{0i0j}^{2}(x,x')\nonumber\\
&=&\left[\frac{|\vec{x}-\vec{x}'|^{2}}{\beta^{2}}\delta_{ij}
-(N-1)\frac{(\vec{x}-\vec{x}')_{i}(\vec{x}-\vec{x}')_{j}}{\beta^{2}}\right]\times\nonumber\\
&&\ \ \ \ \ e^{-2\pi|\vec{x}-\vec{x}'|/\beta}
\cos\left(\frac{2\pi(\tau-\tau')}{\beta}\right)
\frac{\Gamma[(N-1)/2]}{\pi^{(N-3)/2}|\vec{x}-\vec{x}'|^{2N}}\left(\frac{|\vec{x}-\vec{x'}|}{\beta}\right)^{N/2}\times\nonumber\\
&&\ \ \ \ \ \left[1+\frac{(N-2)(N-4)}{16\pi}\left(\frac{\beta}{|\vec{x}-\vec{x}'|}\right)+\cdots\right]+\cdots\\
&&\Delta T_{0ijk}^{2}(x,x')\nonumber\\
&=&e^{-2\pi|\vec{x}-\vec{x}'|/\beta}\sin\left(\frac{2\pi(\tau-\tau')}{\beta}\right)
\frac{\Gamma[(N-1)/2]}{\pi^{(N-3)/2}|\vec{x}-\vec{x}'|^{2N}}\left(\frac{|\vec{x}-\vec{x}'|}{\beta}\right)^{\frac{N}{2}-1}\times\nonumber\\
&&\left\{-(N-2)\frac{|\vec{x}-\vec{x}'|^{2}(\vec{x}-\vec{x}')_{i}}{\beta^{3}}\delta_{jk}
\left[1+\frac{N(N-3)}{32\pi}\left(\frac{\beta}{|\vec{x}-\vec{x}'|}\right)+\cdots\right]\right.\nonumber\\
&&\ \ -\frac{|\vec{x}-\vec{x}'|^{2}\left[\delta_{ij}(\vec{x}-\vec{x}')_{k}+\delta_{ik}(\vec{x}-\vec{x}')_{j}\right]}{\beta^{3}}
\left[1+\frac{N(N-2)}{16\pi}\left(\frac{\beta}{|\vec{x}-\vec{x}'|}\right)+\cdots\right]\nonumber\\
&&\ \ \left.+2(N-1)\frac{(\vec{x}-\vec{x}')_{i}(\vec{x}-\vec{x}')_{j}(\vec{x}-\vec{x}')_{k}}{\beta^{3}}
\left[1+\frac{N(N-2)}{16\pi}\left(\frac{\beta}{|\vec{x}-\vec{x}'|}\right)+\cdots\right]\right\}\nonumber\\
&&+\cdots\\
&&\Delta T_{ijkl}^{2}(x,x')\nonumber\\
&=&\frac{\Gamma^{2}[(N-1)/2]}{8\pi^{N-1}|\vec{x}-\vec{x}'|^{2N}\beta^{2}}\bigg\{
(N^{2}-3N-2)|\vec{x}-\vec{x}'|^{2}\delta_{ij}\delta_{kl}
+2|\vec{x}-\vec{x}'|^{2}\left(\delta_{ik}\delta_{jl}+\delta_{il}\delta_{jk}\right)\nonumber\\
&&\ \ -2(N-1)(N-3)\left[\delta_{ij}(\vec{x}-\vec{x}')_{k}(\vec{x}-\vec{x}')_{l}
+\delta_{kl}(\vec{x}-\vec{x}')_{i}(\vec{x}-\vec{x}')_{j}\right]\nonumber\\
&&\ \ -2(N-1)\left[\delta_{ik}(\vec{x}-\vec{x}')_{j}(\vec{x}-\vec{x}')_{l}
+\delta_{il}(\vec{x}-\vec{x}')_{j}(\vec{x}-\vec{x}')_{k}\right.\nonumber\\
&&\ \ \ \ \ \ \ \ \ \ \ \ \ \ \ \ \left.+\delta_{jk}(\vec{x}-\vec{x}')_{i}(\vec{x}-\vec{x}')_{l}+\delta_{jl}(\vec{x}-\vec{x}')_{i}(\vec{x}-\vec{x}')_{k}\right]\nonumber\\
&&+\frac{4(N-1)^{2}}{|\vec{x}-\vec{x}'|^{2}}
(\vec{x}-\vec{x})_{i}(\vec{x}-\vec{x}')_{j}(\vec{x}-\vec{x}')_{k}(\vec{x}-\vec{x}')_{l}\bigg\}\nonumber\\
&&+e^{-2\pi|\vec{x}-\vec{x}'|/\beta}\cos\left(\frac{2\pi(\tau-\tau')}{\beta}\right)\frac{\Gamma[(N-1)/2]}{\pi^{(N-3)/2}|\vec{x}-\vec{x}'|^{2N}\beta^{2}}
\left(\frac{|\vec{x}-\vec{x}'|}{\beta}\right)^{N/2}\times\nonumber\\
&&\left\{(N-2)|\vec{x}-\vec{x}'|^{2}\delta_{ij}\delta_{kl}
\left[1-\frac{(3N^{2}+6N+32)}{16(N-2)\pi}\left(\frac{\beta}{|\vec{x}-\vec{x}'|}\right)+\cdots\right]\right.\nonumber\\
&&+|\vec{x}-\vec{x}'|^{2}\left(\delta_{ik}\delta_{jl}+\delta_{il}\delta_{jk}\right)
\left[\frac{1}{\pi}\left(\frac{\beta}{|\vec{x}-\vec{x}'|}\right)+\cdots\right]
\nonumber\\
&&\ \ -2(N-2)\left[\delta_{ij}(\vec{x}-\vec{x}')_{k}(\vec{x}-\vec{x}')_{l}
+\delta_{kl}(\vec{x}-\vec{x}')_{i}(\vec{x}-\vec{x}')_{j}\right]\times\nonumber\\
&&\ \ \ \ \ \left[1+\frac{(N^{3}-3N^{2}-26N+16)}{32(N-2)\pi}\left(\frac{\beta}{|\vec{x}-\vec{x}'|}\right)+\dots\right]\nonumber\\
&&\ \ -\left[\delta_{ik}(\vec{x}-\vec{x}')_{j}(\vec{x}-\vec{x}')_{l}
+\delta_{il}(\vec{x}-\vec{x}')_{j}(\vec{x}-\vec{x}')_{k}\right.\nonumber\\
&&\ \ \ \ \ \ \ \ \ \ \ \ \ \ \ \ \left.+\delta_{jk}(\vec{x}-\vec{x}')_{i}(\vec{x}-\vec{x}')_{l}+\delta_{jl}(\vec{x}-\vec{x}')_{i}(\vec{x}-\vec{x}')_{k}\right]\times\nonumber\\
&&\ \ \ \ \ \left[1+\frac{(N^{2}+10N-8)}{16\pi}\left(\frac{\beta}{|\vec{x}-\vec{x}'|}\right)+\cdots\right]\nonumber\\
&&\left.+\frac{4(N-1)}{|\vec{x}-\vec{x}'|^{2}}
(\vec{x}-\vec{x})_{i}(\vec{x}-\vec{x}')_{j}(\vec{x}-\vec{x}')_{k}(\vec{x}-\vec{x}')_{l}
\left[1+\frac{N(N+2)}{16\pi}\left(\frac{\beta}{|\vec{x}-\vec{x}'|}\right)+\cdots\right]\right\}\nonumber\\
&&+\cdots\label{high6}
\end{eqnarray}

\section{Low temperature expansion}\label{appc}

Using Eq.~(\ref{flowt}) we have the low temperature expansions of the correlators

\begin{eqnarray}
\Delta T_{0000}^{2}(x,x')&=&\frac{N\Gamma^{2}(N/2)}{8\pi^{N}\left[|\vec{x}-\vec{x}'|^{2}+(\tau-\tau')^{2}\right]^{N}}\times\nonumber\\
&&\ \ \left[(N-1)-\frac{4N(\tau-\tau')^{2}}{|\vec{x}-\vec{x}'|^{2}+(\tau-\tau')^{2}}
+\frac{4N(\tau-\tau')^{4}}{\left[|\vec{x}-\vec{x}'|^{2}+(\tau-\tau')^{2}\right]^{2}}\right]\nonumber\\
&&-\frac{N\Gamma^{2}(N/2)\zeta(N)}{2\pi^{N}\left[|\vec{x}-\vec{x}'|^{2}+(\tau-\tau')^{2}\right]^{N/2}\beta^{N}}
\left[1-\frac{N(\tau-\tau')^{2}}{|\vec{x}-\vec{x}'|^{2}+(\tau-\tau')^{2}}\right]+\cdots\nonumber\\ \label{low1}\\
\Delta T_{000i}^{2}(x,x')&=&-\frac{N^{2}\Gamma^{2}(N/2)(\tau-\tau')(\vec{x}-\vec{x}')_{i}}{4\pi^{N}
\left[|\vec{x}-\vec{x}'|^{2}+(\tau-\tau')^{2}\right]^{N+1}}\left[1-\frac{2(\tau-\tau')^{2}}{|\vec{x}-\vec{x}'|^{2}+(\tau-\tau')^{2}}\right]\nonumber\\
&&+\frac{N^{2}\Gamma^{2}(N/2)\zeta(N)(\tau-\tau')(\vec{x}-\vec{x}')_{i}}
{2\pi^{N}\left[|\vec{x}-\vec{x}'|^{2}+(\tau-\tau')^{2}\right]^{\frac{N}{2}+1}\beta^{N}}+\cdots
\end{eqnarray}
\begin{eqnarray}
\Delta T_{00ij}^{2}(x,x')&=&\frac{\Gamma^{2}(N/2)}{8\pi^{N}\left[|\vec{x}-\vec{x}'|^{2}+(\tau-\tau')^{2}\right]^{N}}
\left[\delta_{ij}\left((N^{2}-N-4)-\frac{2N(N-2)(\tau-\tau')^{2}}{|\vec{x}-\vec{x}'|^{2}+(\tau-\tau')^{2}}\right)\right.\nonumber\\
&&\ \ \ \ \ \left.-\frac{2N(\vec{x}-\vec{x}')_{i}(\vec{x}-\vec{x}')_{j}}{|\vec{x}-\vec{x}'|^{2}+(\tau-\tau')^{2}}
\left((N-2)-\frac{2N(\tau-\tau')^{2}}{|\vec{x}-\vec{x}'|^{2}+(\tau-\tau')^{2}}\right)\right]\nonumber\\
&&+\frac{\Gamma^{2}(N/2)\zeta(N)}{2\pi^{N}\left[|\vec{x}-\vec{x}'|^{2}+(\tau-\tau')^{2}\right]^{N/2}\beta^{N}}
\left[\delta_{ij}\left((N-4)-\frac{N(N-2)(\tau-\tau')^{2}}{|\vec{x}-\vec{x}'|^{2}+(\tau-\tau')^{2}}\right)\right.\nonumber\\
&&\ \ \ \ \ \ \left.+\frac{2N(\vec{x}-\vec{x}')_{i}(\vec{x}-\vec{x}')_{j}}{|\vec{x}-\vec{x}'|^{2}+(\tau-\tau')^{2}}\right]+\cdots
\end{eqnarray}
\begin{eqnarray}
\Delta T_{0i0j}^{2}(x,x')&=&\frac{\Gamma^{2}(N/2)}{4\pi^{N}\left[|\vec{x}-\vec{x}'|^{2}+(\tau-\tau')^{2}\right]^{N}}
\left[\delta_{ij}\left(1-\frac{N(\tau-\tau')^{2}}{|\vec{x}-\vec{x}'|^{2}+(\tau-\tau')^{2}}\right)\right.\nonumber\\
&&\ \ \ \ \ \left.-\frac{N(\vec{x}-\vec{x}')_{i}(\vec{x}-\vec{x}')_{j}}{|\vec{x}-\vec{x}'|^{2}+(\tau-\tau')^{2}}
\left(1-\frac{2N(\tau-\tau')^{2}}{|\vec{x}-\vec{x}'|^{2}+(\tau-\tau')^{2}}\right)\right]\nonumber\\
&&-\frac{\Gamma^{2}(N/2)\zeta(N)}{2\pi^{N}\left[|\vec{x}-\vec{x}'|^{2}+(\tau-\tau')^{2}\right]^{N/2}\beta^{N}}
\left[\delta_{ij}\left((N-2)+\frac{N(\tau-\tau')^{2}}{|\vec{x}-\vec{x}'|^{2}+(\tau-\tau')^{2}}\right)\right.\nonumber\\
&&\ \ \ \ \ \ \left.-\frac{N(N-1)(\vec{x}-\vec{x}')_{i}(\vec{x}-\vec{x}')_{j}}{|\vec{x}-\vec{x}'|^{2}+(\tau-\tau')^{2}}\right]+\cdots
\end{eqnarray}
\begin{eqnarray}
\Delta T_{0ijk}^{2}(x,x')&=&-\frac{N\Gamma^{2}(N/2)}{4\pi^{N}\left[|\vec{x}-\vec{x}'|^{2}+(\tau-\tau')^{2}\right]^{N}}\times\nonumber\\
&&\left[\frac{(N-2)(\tau-\tau')(\vec{x}-\vec{x}')_{i}\delta_{jk}}{|\vec{x}-\vec{x}'|^{2}+(\tau-\tau')^{2}}
+\frac{(\tau-\tau')(\delta_{ij}(\vec{x}-\vec{x}')_{k}+\delta_{ik}(\vec{x}-\vec{x}')_{j})}{|\vec{x}-\vec{x}'|^{2}+(\tau-\tau')^{2}}\right.\nonumber\\
&&\ \ \ \ \ \left.-\frac{2N(\tau-\tau')(\vec{x}-\vec{x}')_{i}(\vec{x}-\vec{x}')_{j}(\vec{x}-\vec{x}')_{k}}
{\left[|\vec{x}-\vec{x}'|^{2}+(\tau-\tau')^{2}\right]^{2}}\right]\nonumber\\
&&-\frac{N\Gamma^{2}(N/2)\zeta(N)(\tau-\tau')}{2\pi^{N}\left[|\vec{x}-\vec{x}'|^{2}+(\tau-\tau')^{2}\right]^{\frac{N}{2}+1}\beta^{N}}\times\nonumber\\
&&\ \ \left[(N-2)(\vec{x}-\vec{x}')_{i}\delta_{jk}+(\delta_{ij}(\vec{x}-\vec{x}')_{k}+\delta_{ik}(\vec{x}-\vec{x}')_{j})\right]+\cdots
\end{eqnarray}
\begin{eqnarray}
\Delta T_{ijkl}^{2}(x,x')&=&\frac{\Gamma^{2}(N/2)}{2\pi^{N}\left[|\vec{x}-\vec{x}'|^{2}+(\tau-\tau')^{2}\right]^{N}}\times\nonumber\\
&&\left[\frac{N^{2}-N-4}{4}\delta_{ij}\delta_{kl}+\frac{1}{2}\left(\delta_{ik}\delta_{jl}+\delta_{il}\delta_{jk}\right)\right.
\nonumber\\
&&\ \ -\frac{N(N-2)}{2[|\vec{x}-\vec{x}'|^{2}+(\tau-\tau')^{2}]}\left(\delta_{ij}(\vec{x}-\vec{x}')_{k}(\vec{x}-\vec{x}')_{l}
+\delta_{kl}(\vec{x}-\vec{x}')_{i}(\vec{x}-\vec{x}')_{j}\right)\nonumber\\
&&\ \ -\frac{N}{2[|\vec{x}-\vec{x}'|^{2}+(\tau-\tau')^{2}]}
\left[\delta_{ik}(\vec{x}-\vec{x}')_{j}(\vec{x}-\vec{x}')_{l}+\delta_{il}(\vec{x}-\vec{x}')_{j}(\vec{x}-\vec{x}')_{k}\right.\nonumber\\
&&\ \ \ \ \ \left.+\delta_{jk}(\vec{x}-\vec{x}')_{i}(\vec{x}-\vec{x}')_{l}+\delta_{jl}(\vec{x}-\vec{x}')_{i}(\vec{x}-\vec{x}')_{k}\right]\nonumber\\
&&\ \ \left.+\frac{N^{2}}{\left[|\vec{x}-\vec{x}'|^{2}+(\tau-\tau')^{2}\right]^{2}}
(\vec{x}-\vec{x}')_{i}(\vec{x}-\vec{x}')_{j}(\vec{x}-\vec{x}')_{k}(\vec{x}-\vec{x}')_{l}\right]\nonumber\\
&&-\frac{\Gamma^{2}(N/2)\zeta(N)}{2\pi^{N}\left[|\vec{x}-\vec{x}'|^{2}+(\tau-\tau')^{2}\right]^{N/2}\beta^{N}}\times\nonumber\\
&&\left[\delta_{ij}\delta_{kl}\left((N+4)-\frac{N^{2}(\tau-\tau')^{2}}{|\vec{x}-\vec{x}'|^{2}+(\tau-\tau')^{2}}\right)
-2\left(\delta_{ik}\delta_{jl}+\delta_{il}\delta_{jk}\right)
\right.\nonumber\\
&&\ \ +\frac{2N}{|\vec{x}-\vec{x}'|^{2}+(\tau-\tau')^{2}}\left[\delta_{ij}(\vec{x}-\vec{x}')_{k}(\vec{x}-\vec{x}')_{l}
+\delta_{kl}(\vec{x}-\vec{x}')_{i}(\vec{x}-\vec{x}')_{j}\right]\nonumber\\
&&\ \ +\frac{N}{[|\vec{x}-\vec{x}'|^{2}+(\tau-\tau')^{2}]}
\left[\delta_{ik}(\vec{x}-\vec{x}')_{j}(\vec{x}-\vec{x}')_{l}+\delta_{il}(\vec{x}-\vec{x}')_{j}(\vec{x}-\vec{x}')_{k}\right.\nonumber\\
&&\ \ \ \ \ \left.+\delta_{jk}(\vec{x}-\vec{x}')_{i}(\vec{x}-\vec{x}')_{l}+\delta_{jl}(\vec{x}-\vec{x}')_{i}(\vec{x}-\vec{x}')_{k}\right]\bigg]\nonumber\\
&&+\cdots\label{low6}
\end{eqnarray}

\end{document}